\documentclass[11pt,a4paper]{article}
\pdfoutput=1 

\usepackage{jcappub} 
\usepackage[T1]{fontenc} 
\usepackage{bm}

\usepackage{soul}

\usepackage{amsmath}
\usepackage{amssymb}
\usepackage{graphicx}
\usepackage{tabularx}

\usepackage{graphicx}
\usepackage{txfonts}
\usepackage[]{latexsym}
\usepackage[english]{babel}
\usepackage{graphicx}
\usepackage{hyperref}
\usepackage{color}
\usepackage{bm}

\usepackage{tikz}

\usepackage{soul}
\title{\boldmath  Neutrino mass and kinetic gravity braiding degeneracies}

\author[a]{Gabriela Garcia-Arroyo,}
\emailAdd{ gaby.ggarroyo@gmail.com}

\author[b]{Jorge L.~Cervantes-Cota,}
\emailAdd{jorge.cervantes@inin.gob.mx}

\author[c]{Ulises Nucamendi}
\emailAdd{ulises.nucamendi@umich.mx}

\affiliation[a]{Departamento Ingenier\'ia Civil, Divisi\'on de Ingenier\'ia, Universidad de Guanajuato, Av. Ju\'arez No. 77, Guanajuato, C.P. 36000, M\'exico.}

\affiliation[b]{Departamento de F\'isica, Instituto Nacional de Investigaciones Nucleares,
Apartado Postal 18-1027, Col. Escand\'on, Ciudad de M\'exico,11801, M\'exico.}
\affiliation[c]{Instituto de F\'isica y Matem\'aticas, Universidad Michoacana de San Nicol\'as de Hidalgo, Edificio C-3, Ciudad Universitaria, C.P. 58040, Morelia, Michoac\'an, M\'exico.}

\keywords{modified gravity. galaxy clustering. neutrino masses from cosmology}

  \abstract{
  Modified theories of gravity yield an effective dark energy in the background dynamics that achieves an accelerated expansion of the universe. In addition, they present a fifth force that induces gravitational signatures in structure formation, and therefore in the matter power spectrum and related statistics. On the other hand, massive neutrinos suppress the power spectrum at scales that also modified gravity enhances it, so a degeneration of these effects has been recognized for some gravity models.  In the present  work, we study both effects using kinetic gravity braiding ($n$KGB) models to find that in spite of some degeneracies, the role of the fifth force at very large scales imprints a bump in the matter power spectrum as a distinctive signature of this model and, therefore, acts as a smoking gun that seems difficult to match within the present knowledge of power spectra. These models result interesting, however, since the $n=1$ presents no $H_{0}$ tension, and all $n$KGB studied here present no $\sigma_8$ tension and, in addition, a null neutrino mass is excluded.  }

\begin{document} 
\maketitle
\flushbottom


\begin{section}{Introduction}
The standard model of cosmology is based on the existence of dark matter and dark 
energy. The first was initially identified as a necessary ingredient to understand 
the dynamics in galaxies and clusters \cite{Bertone:2004pz}, and later to encompass 
different properties of the Cosmic Microwave Background (CMB), such as its acoustic 
peaks  \cite{Planck:2018vyg}, and to account for the Baryon Acoustic 
Oscillations (BAO) \cite{Zhai:2016gyu}; it is also important for the growth of structure 
in the Universe at different scales. The second is important to achieve an accelerated 
expansion since the last few Giga years \cite{Riess:1998cb,Perlmutter:1998np}, and this 
impacts different observables, such as clustering properties as given by the matter 
power spectrum.  However, none of these dark components have an accepted 
origin so far, hence different approaches have been undertaken.  As a possible alternative 
to dark energy are Modified Gravity (MG) models that apart to generate the accelerated 
expansion, a fifth force associated to their scalar degree of freedom yields distinctive 
cosmological features \cite{Koyama:2015vza}. In particular, the linear cosmic structure 
growth is modified \cite{Aviles:2020wme}, depending now on time and scale, in contrast 
to the one in the $\Lambda$CDM model that is subject only to time evolution.

MG models act mainly in the late universe, when their resulting effective dark energy accelerates the background expansion and its fifth force influences the clustering properties. This in turn has an impact in the CMB statistics at large angles where the cosmic variance makes difficult to distinguish among models; however, Integrated Sachs-Wolfe (ISW) cross correlated with galaxy clustering helps to constrain models \cite{Ferraro:2014msa,Renk:2017rzu,Kable:2021yws}.    On the other hand, structure formation offers a valuable cosmic probe to test gravity on large scales, and therefore to be able to distinguish among different MG models. This can be understood because different fifth forces provoke characteristic effects in the cosmic,  statistical observables, such as the power spectrum (PS) or the two-point correlation function, and general  N-point statistics  \cite{Alam:2020jdv}. However, MG is not the only factor that imprints special features in these statistics. Beyond the $\Lambda$CDM model, massive neutrinos have been recognized to produce a reduction in the structure growth due to its free streaming \cite{Lesgourgues:2006nd} that in turn depends on the sum of their masses;  CMB constraints pose values to less than a fraction of 1 eV, where the exact constraints depend on the data set considered \cite{Planck:2018vyg}. It is then expected that  neutrinos have a range of masses where MG models also yield structure growth effects. But it has been found that MG and massive neutrinos work in opposite directions: whereas neutrinos generate a PS suppression depending on the free streaming scale, MG models cause a PS enhancement due to the attractive fifth force \cite{Bellomo:2016xhl}.  
In recent years the interest to study the joint effects of MG and massive neutrinos has risen, partially because the stage-IV generation of galaxy surveys  and weak lensing probes such as DESI \cite{DESI:2016fyo}, EUCLID \cite{Amendola:2016saw}, or LSST  \cite{LSSTDarkEnergyScience:2012kar} will reach percent-level accuracy in two-point statistics observables, posing the interesting challenge to discriminate among different effects due to gravitational models and neutrino masses.  

The effects of neutrinos in the CMB TT power spectrum have been considered in \cite{Hojjati:2011ix} for the Hu-Sawicki (HS) f(R) gravity model \cite{Hu:2007nk}. It is shown that the presence of neutrinos and a fifth force act in the same direction to lower the anisotropies for low multipoles; higher multipoles are scarcely affected.  But again the cosmic variance prevent us for discriminating these effects.  On the other hand, the effect of massive neutrinos in the PS for different MG models shows that a higher neutrino mass (sum of masses) is allowed in MG to render a similar General Relativity (GR) behavior with a smaller neutrino mass.  This happens because MG rises the matter power spectrum, $P_{\delta \delta}$, whereas the neutrino free streaming lowers it \cite{Bellomo:2016xhl,Hagstotz:2019gsv}; also, a similar effect happens for the weak lensing convergence power spectrum \cite{Peel:2018aly}. Whereas an exact cancellation of these effects does not happen, it seems difficult to distinguish among them. Fortunately, peculiar velocities help to discriminate: the density-velocity and velocity-velocity power spectra, $P_{\delta \theta}$ and  $P_{\theta \theta}$, are different in GR and MG with or without massive neutrinos; in ref. \cite{Hagstotz:2019gsv} this analysis was made for HS model. In the same line of thought and closer to observations, redshift space distortions (RSD) serve to sort out these effects. This has been applied again to HS models in refs.   \cite{Wright:2019qhf,Garcia-Farieta:2019hal}, where analyses are made at different redshifts. 

Other sources of possible parameter degeneracies, such as baryon effects has been recently considered for HS models and CPL dark energy parametrizations \cite{Chevallier:2000qy, Linder:2002et} to reach the levels of accuracy required to test gravity in the above-mentioned probes \cite{Bose:2021mkz}; constraints to the CPL model with massive neutrinos have also been discussed \cite{Zhang:2020mox}. Other studies consider the same HS model to look for degeneracies among neutrinos masses and the halo mass function and the bias \cite{Baldi:2013iza}, however, disentangling the effects is not an easy task, again for the level of accuracy required. Here, voids can help in the analysis \cite{Contarini:2020fdu}. As mentioned, almost all above works considered the HS model. The influence of neutrinos in other gravity models is less common, but in ref. \cite{Barreira:2014ija} the CMB and matter PS for Galileon gravity models are discussed, where again similar results were found, the mass of neutrinos suppresses the enhancement of the fifth force.  

Clustering effects in different gravity models can be different and they have to be studied case by case; another approach is to treat these models through parametrizations \cite{Clifton:2011jh}. The latter approach serves to characterize general features of phenomenological models, whereas the former intends to find consequences of particular models. In the present work we deal with a particular model, a kinetic gravity braiding (KGB) model that is part of the full Horndeski gravity, known to yield second order gravitational field equations. We will consider a  model that is much simpler than the full Horndeski. Its Lagrangian has a kinetic term that depends only on its kinetic energy ($X$), and not on the scalar field itself,  and the coupling to the D'Alembertian of the scalar field, the braiding, also depends only on $X$. This model yields a velocity of light for gravitational wave propagation, so it passes the test of gravity wave speed inferred from the almost simultaneous arrival of light and gravity waves from the merging of neutron stars, among other properties considered in refs.  \cite{Deffayet:2010qz,Kobayashi:2010cm,Kimura:2010di,Arjona:2019rfn}; test against cosmic data for this model are presented in refs.  \cite{Kimura:2011td,Cardona:2020ama}. 

We will work particularly with $n$KGB models \cite{Kimura:2010di} that possess only one or two free parameters, depending on model choices. They  are interesting because within the same setup one obtains the cubic Galileon model for $n=1$ and one recovers the $\Lambda$CDM model at background and first-order perturbation levels  for $n \rightarrow \infty$. This model has two parameters, a power law exponent and its strength. In general, the cosmological background dynamics is not exactly that of $\Lambda$CDM, but since the models have an attractor solution, they will tend to a de Sitter phase, and in fact perturbations will help to distinguish among them.  As it happens in other gravity models that affect large scales, we expect $n$KGB models will also modify linear power spectra, and therefore we find interesting to consider its gravitational effects and those of massive neutrinos, to see if they partially cancel each other, as in the other models mentioned above.

This work is organized as follows: In section \ref{MG_models} we set the theoretical framework of KGB  models, to later understand, in section \ref{degenera}, its effects and degeneracies with massive neutrinos in the cosmological observables. Then, we fit these models to current cosmic data in section \ref{fitting} to understand its role in the $H_0$ and $\sigma_8$ tensions. Finally, our final comments and conclusions are discussed in section \ref{conclu}.       

\end{section}

\begin{section}{Kinetic gravity braiding models} \label{MG_models}

KGB models are the most general, minimally-coupled models   yielding second order field equations. Its general action can be written as \cite{Deffayet:2010qz,Kobayashi:2010cm}:
\begin{eqnarray} 
S_\text{braiding}=\int d^4x\sqrt{-g}\left[\frac{M^{2}_{pl}}{2} R + K(\phi, X) - G(\phi, X) \square \phi + {\cal L}_\text{mat}\right],\label{action}
\end{eqnarray} 
where $R$ is the curvature scalar, $K=K(\phi,X)$ and $G=G(\phi,X)$ are functions of the scalar field $\phi$ and of its kinetic energy density $X\equiv-g^{\mu\nu}(\nabla_{\mu}\phi) (\nabla_{\nu}\phi)/2$, while
$\square\phi\equiv g^{\mu\nu}\nabla_{\mu}\nabla_{\nu}\phi$. The term ${\cal L}_m$ stands for the Lagrangian of the matter degrees of freedom. The gravitational field  equations that can be derived from the above action (\ref{action}) read:
\begin{equation}
 G_{\mu\nu} = \frac{1}{M^{2}_{pl}} \left[ T_{\mu\nu}^{(m)} + T_{\mu\nu}^{(\phi)} \right], 
\label{graveq}
\end{equation}
where $T_{\mu\nu}^{(m)}$ is the total energy momentum tensor, containing dark matter, baryons, neutrinos, and photons, while $T_{\mu\nu}^{(\phi)}$ represents the energy momentum tensor for the Galileon field $\phi$,
\begin{eqnarray} 
T_{\mu\nu}^{(\phi)} = K_{X}(\nabla_{\mu}\phi) (\nabla_{\nu}\phi) + g_{\mu\nu}K + g_{\mu\nu} (\nabla_{\beta}G) (\nabla^{\beta}\phi) 
&& \nonumber \\ 
- \left[(\nabla_{\mu}G) (\nabla_{\nu}\phi) + (\nabla_{\nu}G) (\nabla_{\mu}\phi)\right] - G_{X} (\square \phi) (\nabla_{\mu}\phi) (\nabla_{\nu}\phi)  .
\label{Stress}
\end{eqnarray}

Variation of the action (\ref{action}) with respect to the Galileon scalar field produces the equation,
\begin{eqnarray}
K_{\phi} + \nabla^{\alpha} \left(K_{X} \nabla_{\alpha}\phi\right) + 2G_{\phi\phi}X - 2G_{\phi}\square\phi &&
\nonumber \\
+ G_{X\phi}\left[2X\square\phi + 2(\nabla^{\mu}\phi) (\nabla_{\mu}\nabla_{\nu}\phi) \nabla^{\nu}\phi \right] && \nonumber \\
+ G_{X}\left[ (\nabla_{\mu}\nabla_{\nu}\phi)^2 + R_{\mu\nu} (\nabla^{\mu}\phi)\nabla^{\nu}\phi - (\square\phi)^2 \right] &&
\nonumber \\ 
- G_{XX} \left[ (\nabla_{\nu}X) \nabla^{\nu}X - \nabla^{\mu}\phi (\nabla_{\mu}\nabla_{\nu}\phi) (\nabla^{\nu}\phi) \square \phi \right] & = & 0  \,,
\label{SFE}
\end{eqnarray}
where in eqs. (\ref{Stress}) and (\ref{SFE}) we have defined the first partial derivatives of the coupling functions as,
\begin{equation}
K_{\phi} \equiv \frac{\partial K(\phi,X)}{\partial \phi} , \, 
K_{X} \equiv \frac{\partial K(\phi,X)}{\partial X} , \, 
G_{\phi} \equiv \frac{\partial G(\phi,X)}{\partial \phi} ,  \, 
G_{X} \equiv \frac{\partial G(\phi,X)}{\partial X}  ,
\label{FIRSTDER}
\end{equation}
and the corresponding second partial derivatives as,
\begin{equation}
G_{\phi\phi} \equiv \frac{\partial^{2} G(\phi,X)}{\partial \phi^2} , \quad
G_{X\phi} \equiv \frac{\partial^{2} G(\phi,X)}{\partial X \partial \phi} , \quad
G_{XX} \equiv \frac{\partial^{2} G(\phi,X)}{\partial X^{2}} . 
\label{SECONDDER}
\end{equation}
We note that in the action (\ref{action}) the interaction term $G\square\phi$ contains the coupling of $G$ with first-order derivatives of both the metric and scalar field  stemming from the Christoffel symbols contained in the D'Alembertian operator. These couplings produce a class of mixing named kinetic braiding, in which the essential characteristic is determined by the kinetic dependence of $G=G(X)$. This implies that the scalar field equation of motion and its energy momentum tensor contain second derivatives of the metric in an essential way: there exists no Einstein frame where the kinetic terms are diagonalized \cite{Deffayet:2010qz}. There are many possibilities to choose the $K$ and $G$ functions, we chose $G = G(X)$ because is precisely the coupling of $X$ with $\square\phi$ that is responsible of the braiding mixing; we also chose the kinetic term as a function of the kinetic energy only, $K=K(X)$, to avoid extra complexities, as it can be seen in equation (\ref{SFE}). As a consequence, the action (\ref{action}) is invariant under the change 
$\phi \longrightarrow \phi + b$ (where b is a constant); this shift symmetry gives a conserved Noether current $J_{\mu}$ given by, 
\begin{equation}
 J_{\mu} = \left( K_{X} - G_{X} \square \phi \right) \nabla_{\mu}\phi + G_{X} (\nabla_{\mu}\nabla_{\nu}\phi) \nabla^{\nu}\phi \, .
 \label{NOETHER}
\end{equation}
The conservation equation for the Noether current $\nabla_{\mu}J^{\mu} = 0$ gives directly the Galileon field equation (\ref{SFE}) with $K_{\phi} = G_{\phi}= 0$,
\begin{eqnarray}
&&\nabla^{\alpha} \left(K_{X} \nabla_{\alpha}\phi\right) 
+ G_{X}\left[ (\nabla_{\mu}\nabla_{\nu}\phi)^2 + R_{\mu\nu} (\nabla^{\mu}\phi)\nabla^{\nu}\phi - (\square\phi)^2 \right] 
\nonumber \\ 
&& - G_{XX} \left[ (\nabla_{\nu}X) \nabla^{\nu}X - \nabla^{\mu}\phi (\nabla_{\mu}\nabla_{\nu}\phi) (\nabla^{\nu}\phi) \square \phi \right] =  0  \, .
\label{Galileon_equation}
\end{eqnarray}

Before we specify the $K(X)$ and $G(X)$ functions we would like to show the Friedmann equations. 
For the spatially flat Friedmann-Robertson-Walker metric we have: 
\begin{eqnarray}\label{friedmannEQ}
3 H^{2} &=& (8\pi G) \left[ \rho_{m} + \rho_{r} + \rho_{\phi} \right] \quad ,  \\
2\dot{H} + 3H^{2} &=& - (8\pi G)  \left[ P_{r} + P_{\phi} \right] \quad ,
\end{eqnarray}
where the dot denotes the derivative with respect to the cosmic time, $H$ is the Hubble parameter, and the matter and radiation densities obey standard fluid equations.  The density and pressure of the Galileon field are given by:
\begin{eqnarray}\label{galileonDP}
\rho_{\phi} &=& -K + K_{X} \dot{\phi}^{2} + 3G_{X} H \dot{\phi}^{3}  \quad, \\
P_{\phi} &=& K - G_{X}\dot{\phi}^{2} \ddot{\phi}  \quad .
\end{eqnarray}

The scalar field equation (\ref{Galileon_equation}) becomes, 
\begin{eqnarray}
K_{X} \left( \ddot{\phi} + 3H\dot{\phi} \right) 
+ K_{XX}  \ddot{\phi} \dot{\phi}^{2} 
+ 3G_{X} \left( 2H\dot{\phi}\ddot{\phi} + 3H^{2}\dot{\phi}^{2} + \dot{H}\dot{\phi}^{2} \right)
+ 3G_{XX}H \dot{\phi}^{3} \ddot{\phi} & = & 0  \,.
\label{GEFRW}
\end{eqnarray}



In the rest of this work we consider the $n$KGB model \cite{Kimura:2010di}: 
\begin{eqnarray} \label{K}
K(X) &=& -X  \quad, \\
G(X) &=& g^{(2n-1)/2} \, \Lambda \left(\frac{X}{\Lambda^4} \right)^{n}    \quad ,  \label{G} 
\end{eqnarray}
where $g$ is a dimensionless constant that controls the strength of the braiding and $\Lambda$ has dimensions of energy weighted by the Hubble constant and the Planck mass such that $\Lambda^{4n-1} = H_{0}^{2n} M_{pl}^{2n-1}$ \cite{Bellini:2019syt}. Note that the l.h.s. of this later equation should be provided by a particle physics model, the pair $(\Lambda, n)$, whereas the r.h.s. is fixed by cosmology.  As $n$ tends to infinity, the cosmological constant scale is $\Lambda = \sqrt{H_{0} \, M_{pl}} \approx 10^{-3} eV$, as in the $\Lambda CDM$ model, assuming a Hubble constant $H_0$ that matches observations. For smaller $n$, the scale of $\Lambda$ diminishes since it is suppressed by $n-$powers of the Planck's constant,  say for $n=3$,  $\Lambda = (H_{0}^{6} \, M_{pl}^{5})^{1/11} \approx 10^{-6} eV$ and for $n=1$,  $\Lambda = (H_{0}^{2} \, M_{pl})^{1/3} \approx 10^{-13} eV$.  Then, smaller $n$ aggravates the cosmological constant problem. However, from other theoretical and cosmological grounds such models could be of interest, see e.g.  \cite{Deffayet:2010qz}. For example, the election $n=1$ corresponds to the covariant, cubic  galileon, that is in big tension with ISW cross correlated to clustering data \cite{Renk:2017rzu}, but we included it here for completeness.

$n$KGB models are interesting because the sound speed for scalar perturbations can be smaller or bigger than light velocity, see Eqs. (A.11)-(A.12) and subsequent paragraph of \cite{Deffayet:2010qz}, hence, its clustering properties are nontrivial; another important characteristic is the existence of an accelerated attractor exhibiting phantom behaviour in its evolution that tends to a de Sitter phase, as we will discuss below.  In attractor phase and for models with $n > 1/2$, it has been proven that this model does not have ghosts and instabilities for the cosmological perturbations \cite{Deffayet:2010qz,Kimura:2011td}. Important for our work, their background and linear cosmological perturbations tend to the $\Lambda$CDM model for $n \rightarrow \infty$. Also, the sound speed of the tensor modes is equal to that of light since it is not affected by the braiding coupling, see the related discussion in Subsection 4.2 of \cite{Kobayashi:2019hrl}.  Additionally, these models have a well-posed initial value problem  and have a Vainshtein mechanism that recover GR at Solar System scales \cite{Deffayet:2010qz}.

In the flat FRW spacetime, the conserved charge density of the Noether current (\ref{NOETHER}) becomes 
\begin{equation} \label{Noether current}
    J_{0} = \dot{\phi} (3 \dot{\phi} G_{X}H - 1) \, .
\end{equation}

It has been shown in ref. \cite{Deffayet:2010qz} that the simplest attractor solutions are found under the condition $J_0=0$. From (\ref{Noether current}) we have two branches,
\begin{eqnarray}\label{attractor1}
\dot{\phi}&=&0  \, , \\ 
\dot{\phi}&=& \frac{1}{3G_{X}H}   \, ,
\label{attractor2}
\end{eqnarray}
the branch $\dot{\phi}=0$ develops ghosts in its cosmological perturbations \cite{Kimura:2011td} and it is not of physical interest. The second branch (\ref{attractor2}) is self-accelerating and well behaved under cosmological perturbations even though its  equation of state is of phantom type.  Let's mention that an extensive analysis of scalar cosmological perturbations (without the inclusion of massive neutrinos) around this background self-accelerating attractor has been done in \cite{Kimura:2011td}. An important property is that its asymptotic evolution describes a de Sitter attractor that can be analysed through the total equation of state 
\begin{equation} \label{w-attractor}
  w = -1 -\frac{2}{3}\frac{\dot H}{H^2} = -1 + \frac{(2n-1)}{3}\frac{(3\Omega_{m} + 4 \Omega_{r})}{(2n-\Omega_m-\Omega_{r})} \,,  
\end{equation}
where the $\Omega_i$ are functions of time. This expression encodes the effective evolution of the background dynamics.  Evaluated  today, $w=-0.82$ for $n=1$ and $w=-0.7$ for  $n \rightarrow \infty$.   In this work, we do not force the background evolution to the self-accelerating attractor, but we take an arbitrary background setup. However, for some of models  we will recover the attractor solution, when parameters fit to cosmological data, as  explained below.



We now proceed to briefly describe the field equations for the linear cosmological perturbations of the corresponding $n$KGB model around the FRW background spacetime. The perturbed metric line element is written in the Newtonian gauge as
\begin{equation}
    ds^{2} = -(1+2\Psi) dt^{2} + a^2(t) (1+2\Phi) \delta_{ij} dx^{i}dx^{j}  \quad,
    \label{Pmetric}
\end{equation}
while the scalar field is expanded by the background solution $\phi(t)$ and its perturbation $\delta\phi$ as 
\begin{equation}
    \phi(x^{i},t) =  \phi(t) + \delta\phi(x^{i},t)
    \label{PSF} \, .
\end{equation}
For the perturbations of the energy momentum tensor we have,
\begin{eqnarray}
\delta T^{i}_{\,\,j} &=& \delta^{i}_{j} (\delta p) + (\partial_{i}\partial_{j} \Pi) \,, \\
\delta T^{i}_{\,\,t} &=& - \frac{1}{a^2} \partial_{i}(\delta q) \,, \\
\delta T^{t}_{\,\,i} &=& \partial_{i}(\delta q) \,, \\
\delta T^{t}_{\,\,t} &=& - (\delta \rho) \,, 
\label{EMTperturbed}
\end{eqnarray}
where $\delta \rho$, $\delta p$, $\delta q$ denote the perturbations of the  total energy density, pressure, and  velocity field, respectively, while $\Pi$ describes the total anisotropic stress.  

The perturbed cosmological Einstein equations are given by equations (4.2), (4.4), (4.6) of reference  \cite{Kimura:2010di} together with the equation (4.5) modified here to incorporate the anisotropic stress $\Pi$,
\begin{equation}
M^{2}_{pl} (\Psi + \Phi) = -a^{2} \Pi  \,,
\label{equationij}
\end{equation}
while the scalar field equation is described by expression (4.7) in \cite{Kimura:2010di} and it will not be rewritten here. Note that in this work $\Pi \neq 0$, though the gravitational part has no slip, we consider massive neutrinos that have a small anisotropic stress $\Pi_{\nu}$, 
apart from those of photons.

Finally, the perturbed energy momentum conservation equations for every matter component are,
\begin{equation}
    \dot{\delta \rho_{i}} + 3H \left[ \delta\rho_{i} + \delta p_{i} + \frac{1}{3} \nabla^{2} \Pi_{i} \right] + 3(\rho_{i} + p_{i}) \dot{\Phi} + \frac{1}{a^2}\nabla^{2} (\delta q_{i}) = 0   \,,
    \label{EC}
\end{equation}
\begin{equation}
 \dot{\delta q_{i}} + 3H\delta q_{i} + (\rho_{i} + p_{i}) \Psi + \delta p_{i} + \nabla^{2} \Pi_{i} = 0 \,,
 \label{MC}
\end{equation}
\\
where $i$ denotes CDM, baryons, radiation, and neutrinos.

%

%
\end{section}

\begin{section}{MG and neutrino mass degeneracies} \label{degenera}
The effects of massive neutrinos on the CMB temperature power spectrum are four \cite{ParticleDataGroup:2020ssz}: 
i) A change in the  nonrelativistic energy density induces changes between scales on the last scattering surface and angles on the sky that affect the angular diameter distance to last scattering. And, on the other hand, it changes the late ISW effect, that depends on the redshift at  matter-to-DE equality; 
ii) variation of the metric fluctuations when they become nonrelativistic; iii) large neutrino masses tend to affect less the CMB photons due to LSS  weak lensing; iv) neutrinos that become nonrelativistic earlier, due to smaller momenta, are imprinted in high-$\ell$ multipoles.  The first and the third effects are more important to put bounds on the masses. On the other hand, MG changes the metric potentials involved in the evolution of the perturbations at large scales and therefore changes the late ISW effect and the clustering properties of matter that induces differences in the lensing properties.


Here we show results for $n$KGB models, using hi\_class \cite{Zumalacarregui:2016pph} to compute the spectrum for different $n$ powers, see eqs. (\ref{K}) and (\ref{G}), where we assumed three species of neutrino with degenerate mass, leaving all other parameters  fixed; for comparison we assume a Planck base-$\Lambda$CDM model with $M_\nu \equiv \Sigma m_{\nu} = 0.06$eV. In figure \ref{fig:cmb_tt} one observes a larger neutrino mass lowers low multipoles, but braiding gravity enhances them; a similar behavior is observed for the $n=1$ model in \cite{Renk:2017rzu}. Due to the point ii) above, a dip is observed at around $\ell \sim 40-200$ (for the $\Lambda$CDM version see \cite{ParticleDataGroup:2020ssz}), that here has two counter effects: whereas massive neutrinos slightly change potentials wells when they become nonrelativistic, the gravity effect works in the other direction at these scales; this latter effect is more prominent for smaller $n$.  The oscillatory pattern for $\ell > 200$,  known in massive neutrino $\Lambda$CDM cosmologies, see point iii) above,  is also present here, the more pronounced for smaller $n$.  
One observes  that the $n=3$ curve is closer to $\Lambda$CDM  than the $n=1$ curve, since the time variation of gravitational potentials in $n$KGB are larger for smaller $n$ \cite{Kimura:2010di,Kimura:2011td}, and  tends to GR for $n \rightarrow \infty$. Particularly this effect is seen for modes  $\ell < 30$, the ISW region.  
In other MG theories the effect induced by the gravity model could be different to the presented in this work, for example, in ref. \cite{Hojjati:2011ix} results are shown for the HS model and the deviations of the CMB TT  angular power spectrum looks different from figure \ref{fig:cmb_tt}, but the effect caused by the sum of neutrino mass is the same.

\begin{figure}[ht]
     \centering
     \includegraphics[width=.7\textwidth]{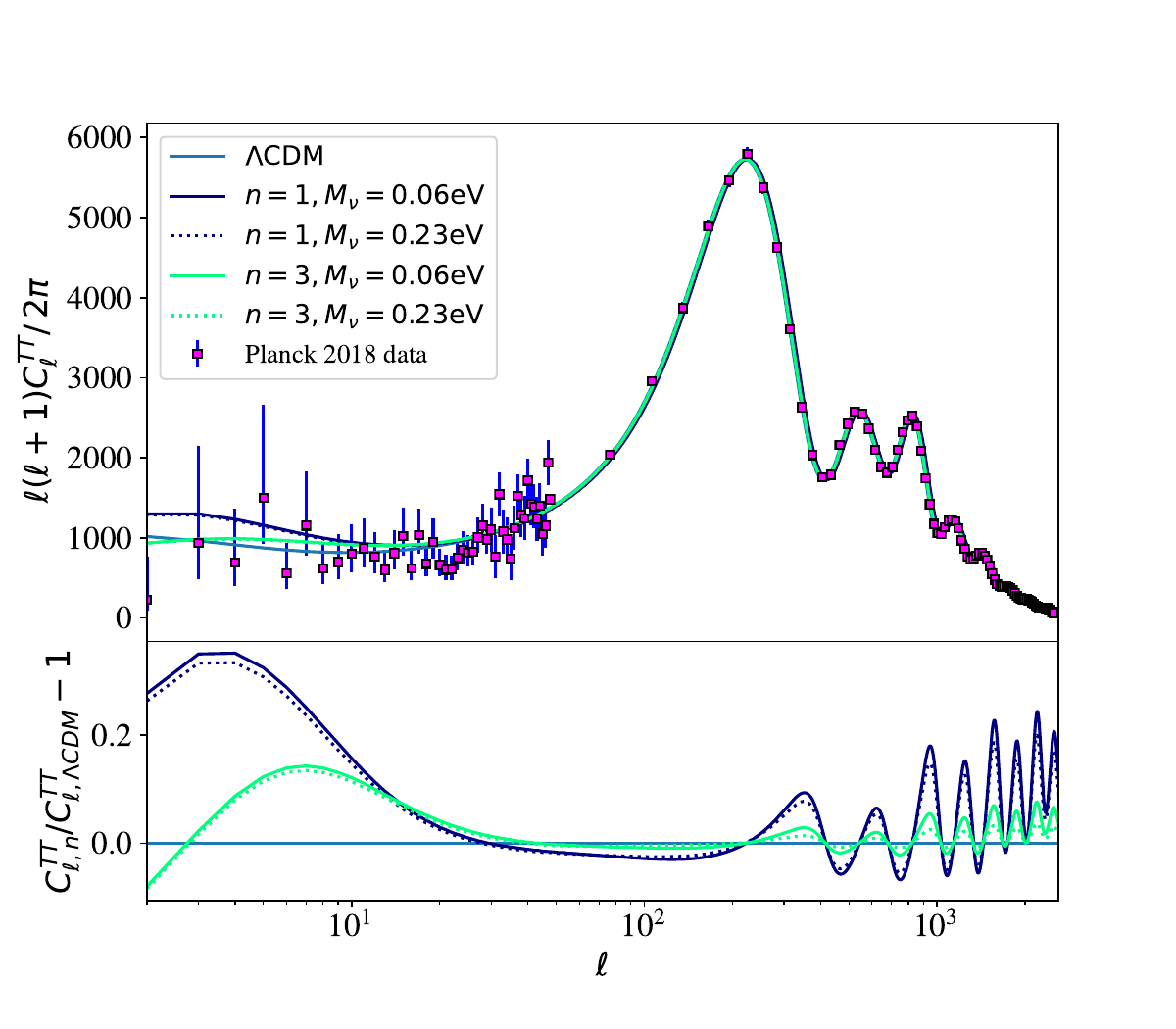}
     \caption{CMB TT power spectrum for $n$KGB models  with $n=1, 3$  and different sum of neutrino masses $M_{\nu}$. Solid lines correspond to $M_{\nu}=0.06 \, \rm{eV}$ and dotted lines to $M_{\nu}=0.23 \,  \rm{eV}$, where we assumed three species of neutrino with degenerate mass. Colors denote different $n$ values. Data points are those from Planck 2018 \cite{Planck:2018vyg}. The bottom panel shows the relative difference to $\Lambda$CDM.  }
     \label{fig:cmb_tt}
 \end{figure}
 
We now look for effects on large scale structure observables. Different combinations of cosmological parameters can cause similar outcomes in an observable, and this may prevent us to distinguish which parameter is responsible for the observed effects.  In the case of massive neutrinos, it is well known that they produce a decrease in dark matter perturbations which in turn is manifested in the matter PS that presents a reduction  \cite{Lesgourgues:2006nd} at wavenumbers ($k$) between $0.01$ and $1$ $h$ Mpc$^{-1}$ depending on their masses. On the other hand, modifications of  gravity can also alter the PS, since the fifth force involved increases the gravitational potential, and the PS grows.     

These two effects are shown in figure \ref{fig:mps} using linear power spectra at $z=1$ (left panel) and $z=0$ (right panel) for the $n$KGB models ($n=1,3$),  discussed in section \ref{MG_models} employing equations (\ref{K}, \ref{G}), where the value of $g$ is fixed by the attractor solution to $0.027$ ($n=1$) and $0.272$ ($n=3$), and $\Omega_m=0.316$ in both cases as the best fit of CMB TT,TE,EE+lowE+lensing from Planck 2018 \cite{Planck:2018vyg}. Here we plot the relative matter PS with respect our base $\Lambda$CDM model.

As in other gravity models, the enhancement produced due to the fifth force is reduced by the total neutrino mass ($M_\nu$).  The bigger $M_\nu$, the streaming effect grows, resulting in a more decreased PS.  It has been shown in HS models \cite{Bellomo:2016xhl} and in Galileon \cite{Barreira:2014ija} that one cannot distinguish these models with a bigger neutrino from GR with a smaller neutrino mass. The effects are not exactly cancelled out, but the distinction among models  results unpractical for the level of precision we have nowadays or in the near future; DESI will be providing PS determinations within 1\% \cite{DESI:2016fyo}.  
Up to intermediate wavenumbers ($k \gtrsim 0.1$ $h$/Mpc) deviations can be hidden with a selected $M_\nu$ (left panel), that in our case we found $M_\nu = 0.14$ eV, to produce tiny differences of less than 1\%. However, the bump at linear scales, $k \sim 0.005$ Mpc/$h$ represents a big deviation (28\% for $n=1$ and 25 \% for $n=3$) from $\Lambda$CDM; generic bumps in the PS have been recently studied in ref. \cite{Gomez-Navarro:2020fef}, where it is shown that a second smaller bump is induced at nonlinear scales. As the universe evolves, clustering increases and at $z=0$ (right panel), at wavenumbers $k \gtrsim 0.1$ $h$/Mpc the differences are bigger than 5\% for $M_\nu =0.23$ eV, and they can be larger for smaller  neutrino masses. To again force the differences to tiny values (less than 1\%) at $z=0$ for these scales, taking the $n=3$ model (that is the one closest to GR in our analysis), one would need a neutrino mass as big as $M_\nu = 0.34$ eV, or $M_\nu = 0.36$ eV for negative relative PS, values that are still compatible to terrestrial neutrino mass measurements. We  show this in figure \ref{fig:mps2}, but the bump at $k \sim 0.01$ $h$/Mpc is still there and represents a big deviation (15 \%) from $\Lambda$CDM. Therefore, this model 
cannot get mixed up with GR provided with lower masses.

Other gravity models such as HS  \cite{Bellomo:2016xhl,Hagstotz:2019gsv} or Galileon \cite{Barreira:2014ija} for some high range of neutrino masses are degenerate with GR assuming low masses at the level of precision required for the matter PS. As demonstrated above, in $n$KGB models a distinction is possible even for the matter PS. However, more features are expected, and further ways to distinguish gravity models, when one considers peculiar velocities. Then, apart from linear overdensity statistics, $P_{\delta \delta} \equiv  P_L(k) $, it is useful to consider the linear density-velocity power spectrum, $P_{\delta \theta} =  \frac{f(k)}{f_0} P_L(k) $, and the velocity-velocity spectrum, $ P_{\theta\theta} =  \left(\frac{f(k)}{f_0} \right)^2 P_L(k)$, where $f_{0}(z) = f(k=0,z)$ is the growth rate in $\Lambda$CDM that is independent of the wavenumber, for further details we refer to \cite{Aviles:2020wme}. In ref.  \cite{Hagstotz:2019gsv} is shown  that the power spectra $P_{\delta \theta}$ and  $P_{\theta \theta}$ are different for the HS model with massive neutrinos from GR (with or) without massive neutrinos. In the same line of thought, RSD serve to sort out these effects for HS  models \cite{Wright:2019qhf,Garcia-Farieta:2019hal}. The main ingredient lies in the growth rate, that for our models depends on scale and time: 

\begin{equation} \label{growth_rate}
f(k,z) \equiv \frac{d {\rm ln} D_+}{d {\rm ln} a} = -\frac{1}{2} \,  \frac{(1+z)}{P_L} \, \frac{dP_L}{dz}\, ,
\end{equation} 
where $D_{+} (k, z)$ is the growth function. Using eq.  (\ref{growth_rate}) we computed $f(k, z)$ for $n=1$ and $n=3$ KGB models that are shown in figure \ref{fig:f_k_z} evaluated at $z=1$ (left panel) and $z=0$ (right panel). 
At large scales (small $k$) the growth rates of models with different neutrinos masses tend to the same value, since there  massive neutrinos behave simply as cold dark matter. At $z=1$,  it grows around 4\% between $k=10^{-3}$ and $k=10^{-2}$ $h$/Mpc for the $n=3$ model and 8\% for $n=1$.
At $z=0$, changes are more evident, it grows around 35\% between $k=10^{-3}$ and $k=10^{-2}$ $h$/Mpc for the $n=3$ model and 40\% for $n=1$. This behavior is expected since the fifth force, that enhances the gravitational pull, increases the growth of structures, as well.  Smaller  deviations of the growth rate are reported for the HS  model \cite{Li:2015poa} at $z=0$. We also plot in figure \ref{fig:croos_power} the resulting $P_{\delta \theta}$ and $P_{\theta \theta}$ with respect to $\Lambda$CDM power spectra, where the differences to the later model are evident. The most important effect is in $P_{\theta \theta}$ since it depends quadratically on $f(k,z)$ in comparison to $P_{\delta \theta}$ that depends on it linearly. Overall the effect of neutrino mass in $f(k,z)$ is much less important than $n$KGB gravitation.       

Nonlinear deviations are computed in ref. \cite{Aviles:2020wme} for the HS model with zero neutrino mass, where it is shown that nonlinear terms tend to decrease $P_{\delta \theta}$ and $P_{\theta \theta}$ in the quasilinear regime. Massive neutrinos have been recently considered nonlinearly in RSD at 1-loop using Eulerian perturbation theory for $\Lambda$CDM model \cite{Aviles:2021que}, resulting in an excellent agreement with simulation data up to $k \sim 0.25$ $h$/Mpc.

 \begin{figure}[ht]
     \centering
     \includegraphics[width=.48\textwidth]{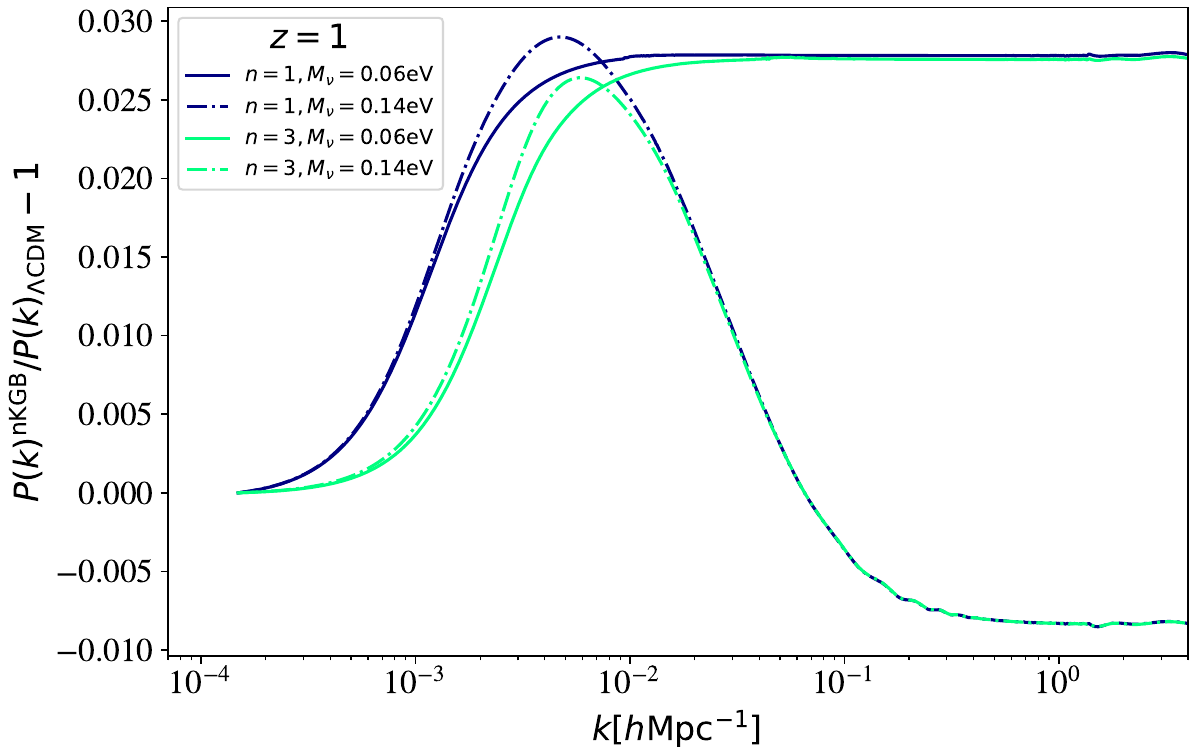}
     \includegraphics[width=.48\textwidth]{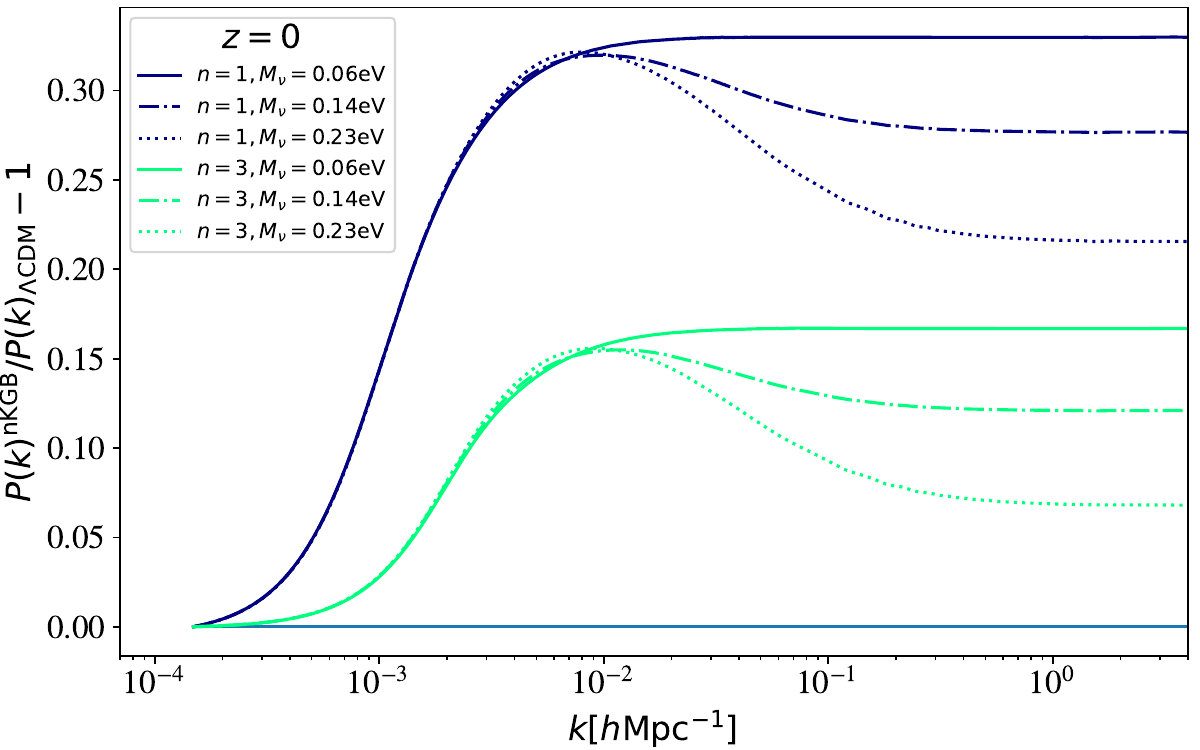}
     \caption{Relative change of the matter PS for $n$KGB models with respect to the one for the $\Lambda$CDM model. The left panel is computed at redshift $z=1$ while the right panel at $z=0$.  Solid lines correspond to $M_{\nu}=0.06 \rm{eV}$, dash--dotted to  $M_{\nu}=0.14 \,  \rm{eV}$, and dotted lines to $M_{\nu}=0.23 \, \rm{eV}$; colors denote different $n$KGB models. }
     \label{fig:mps}
 \end{figure}

 \begin{figure}[ht]
     \centering
     \includegraphics[width=.68\textwidth]{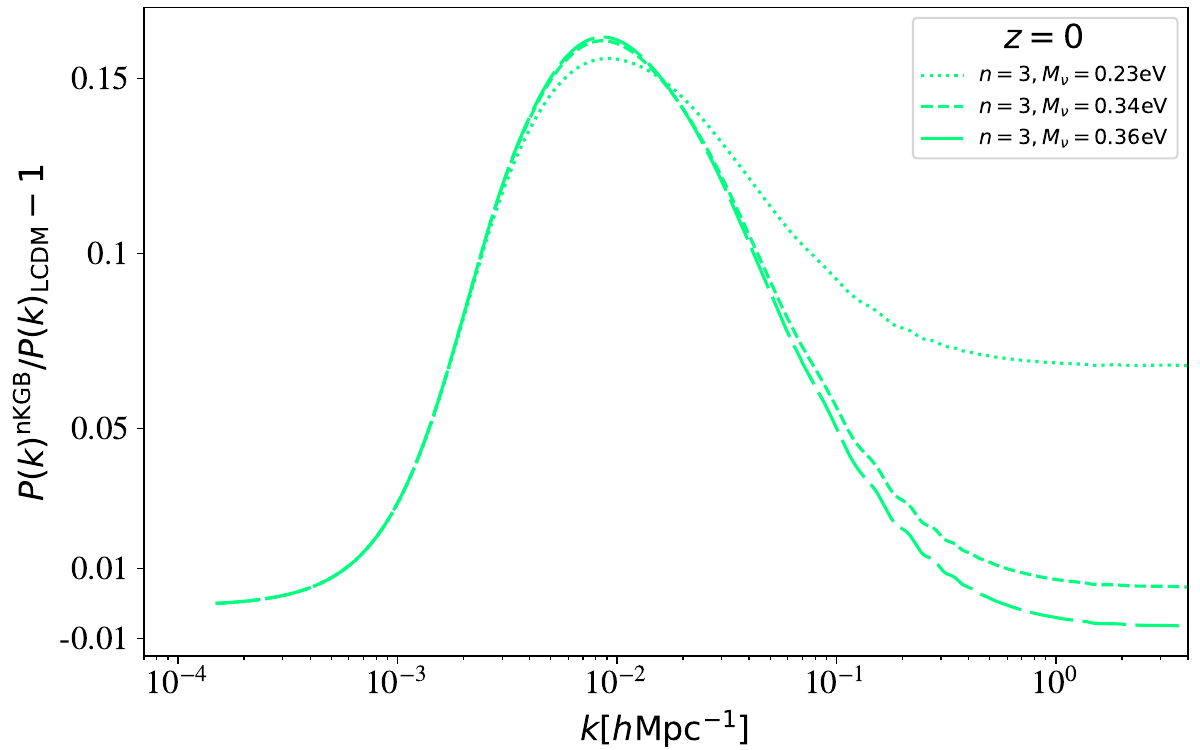}
     \caption{Relative change of the matter PS for the $n=3$ KGB model with respect to the one for the $\Lambda$CDM model  at redshift $z=0$.  Dotted lines correspond to $M_{\nu}=0.23  \rm{eV}$, while small-dashed are for $M_{\nu}=0.34 \,  \rm{eV}$ that yields a 1\% PS difference and long-dashed lines correspond to $M_{\nu}=0.36 \, \rm{eV}$ that yields a -1\% PS difference. A bump at $k \sim 0.01$ $h$/Mpc is the smoking gun of this model.}
     \label{fig:mps2}
 \end{figure}
  
  \begin{figure}[ht]
     \centering
     \includegraphics[width=.48\textwidth]{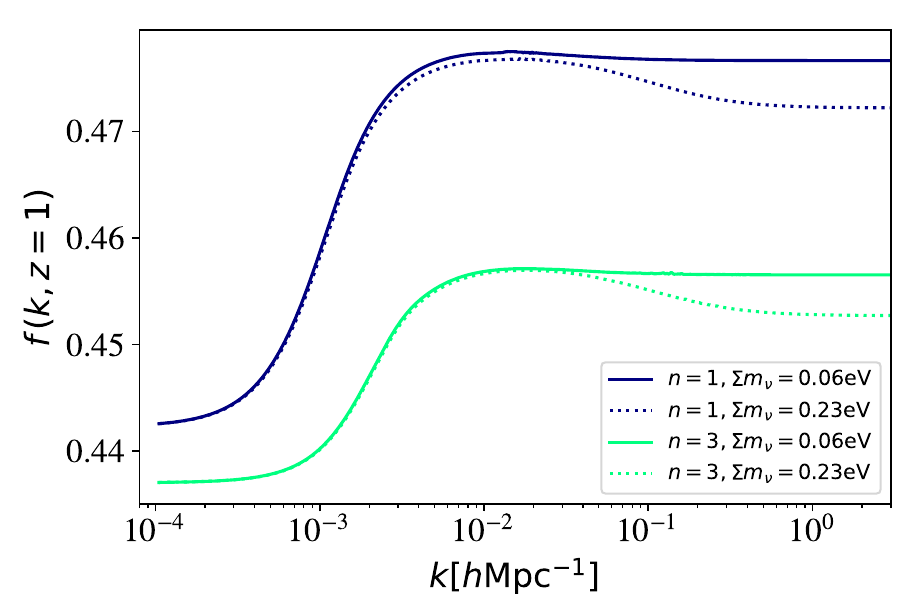}
     \includegraphics[width=.48\textwidth]{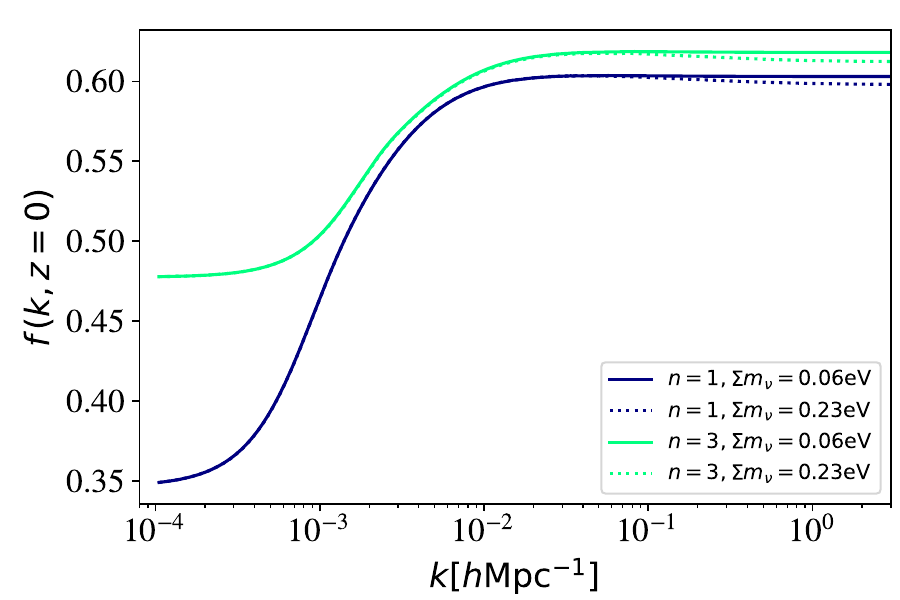}
     \caption{Scale--dependent growth rate for $n$KGB models and different sum of neutrino masses. Left panel is computed at $z=1$ while the right panel is evaluated at $z=0$.}
     \label{fig:f_k_z}
 \end{figure}
 \begin{figure}[ht]
     \centering
     \includegraphics[width=.48\textwidth]{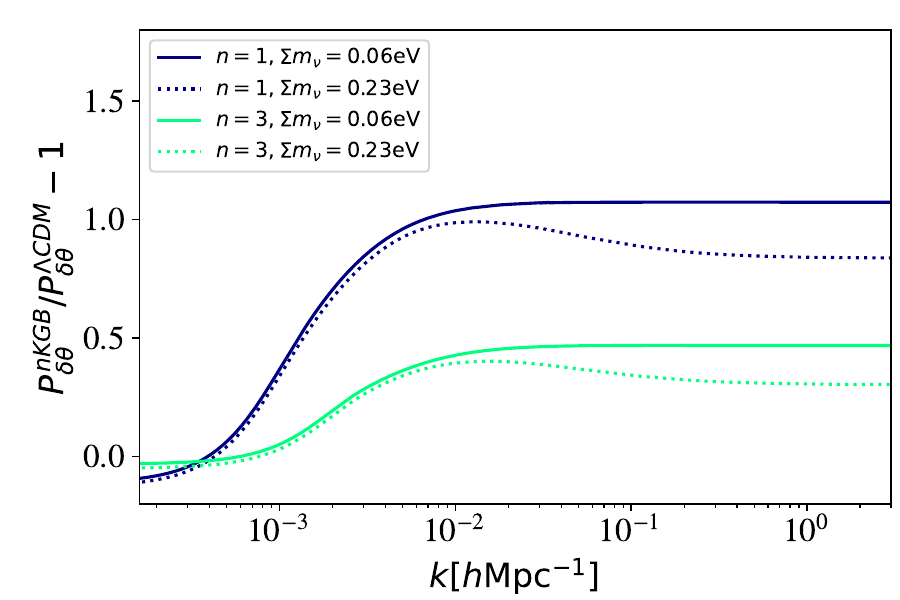}
     \includegraphics[width=.48\textwidth]{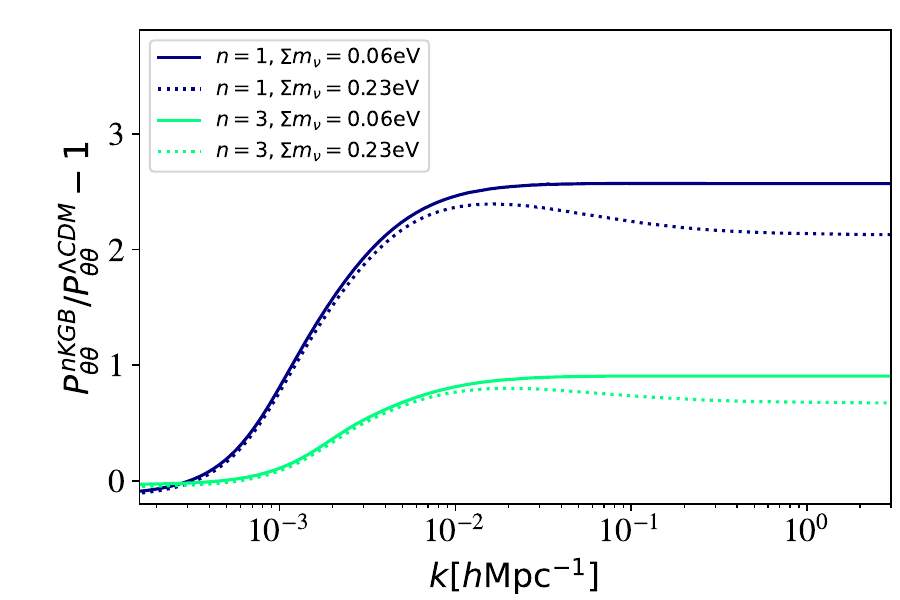}
     \caption{Relative change of linear power spectra for $n$KGB models and different sum of neutrino masses evaluated at $z=0$. The left panel shows the cross-spectrum between the overdensity $\delta$ and the velocity divergence $\theta$. The right panel is for the velocity divergence PS.}
     \label{fig:croos_power}
 \end{figure}
 
\end{section}

\begin{section}{Fitting nKGB models to cosmological data} \label{fitting}

We will constrain the base cosmological parameters, the sum of neutrino masses as well as the parameters of $n$KGB models. This MG model has in principle two extra parameters, $g$ and $n$ in eq. (\ref{G}). $g$ controls  the strength of the braiding and $n$ is the exponent of the kinetic coupling. In order to disentangle both contributions, first we chose to explore a free amplitude $g$, at fixed  $n$--values. We consider separately models with $n=1,2,$ and $3$. And we know that as $n \rightarrow \infty$, the models tend to $\Lambda$CDM. Later, we will consider both $g$ and $n$ free parameters.

We use the Einstein--Boltzmann solver hi\_class \cite{Zumalacarregui:2016pph}, which incorporates the field equations presented in section \ref{MG_models}, and the Markov chain Monte Carlo code  MontePython \cite{Brinckmann:2018cvx} to sample the parameters. For the statistical analysis we employ the following data sets (Planck + BAO):  Planck 2018 TT,TE,EE+lowE+lensing \cite{Planck:2018vyg} and BAO BOSS DR12 \cite{BOSS:2016wmc}.  We include Planck lensing data, because there is a known dependency between neutrino mass and lensing, as explained in point iii) of section \ref{degenera}; see however possible issues with lensing data from Planck and other probes  \cite{DiValentino:2021imh}. Since massive neutrinos are thought to become  nonrelativistic after last scattering, they change the angular diameter distance to this surface, point i) of section \ref{degenera}, and therefore the neutrino mass ($M_\nu$) is  degenerate with $H_0$. To break this parameter degeneracy, it is important to include BAO data (that is measured at low z), to fit better the acoustic scale.  On the other side, we also use correlation function measurements of the KiDS collaboration  \cite{Hildebrandt:2016iqg,Kohlinger:2017sxk} set alone to find out the level of the $\sigma_8$-tension between weak lensing and Planck+BAO data in these MG models.As supernovae data have less constraining power for neutrino masses, we have not included them here.

The contour confidence plots at $68\%$ and $95\%$ for the parameters of interest are shown in figures \ref{fig:n1_1}, \ref{fig:n1_2},  and \ref{fig:n1_3} for the $n=1,2,3$ KGB models, respectively, and the best fit results are reported in table \ref{table1} at $95 \%$ confidence level (CL.) for the varied parameters $(\omega_{b}, \omega_{c}, H_{0}, A_{s}, n_{s} , \tau, M_{\nu}, g)$ and derived ones $(\Omega_{m}, \sigma_{8})$. Neutrinos masses are given in $eV$.

\begin{figure}
    \centering
    \includegraphics[width=.48\textwidth]{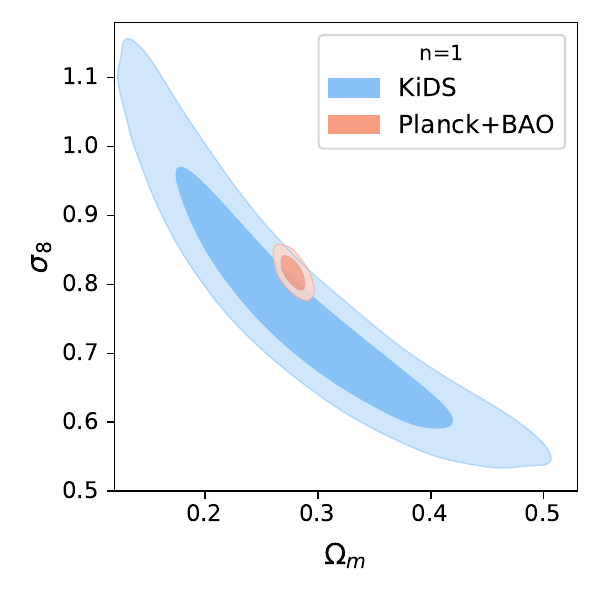}
    \includegraphics[width=.48\textwidth]{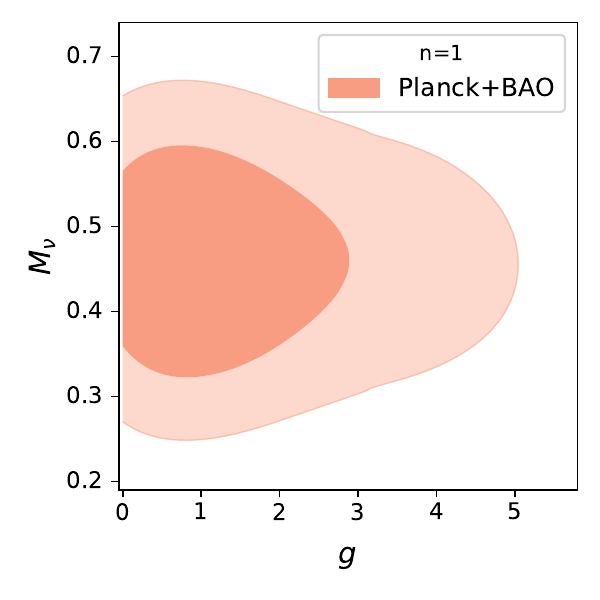} 
    \caption{Left panel: Contour confidence plots at $68\%$ and $95\%$ CL. of $\sigma_8$ vs $\Omega_m$ for the $n=1$ KGB model using Planck+ BAO and KiDS data; in this model the $\sigma_8$ tension between these data is around $0.28\sigma$, so it essentially disappears. Right panel: $M_\nu$ [$eV$] vs $g$ for the Planck+ BAO data; we did not include KiDS data since this dataset has much less constraining power for these parameters.}
    \label{fig:n1_1}
\end{figure}

\begin{figure}
    \centering
    \includegraphics[width=.48\textwidth]{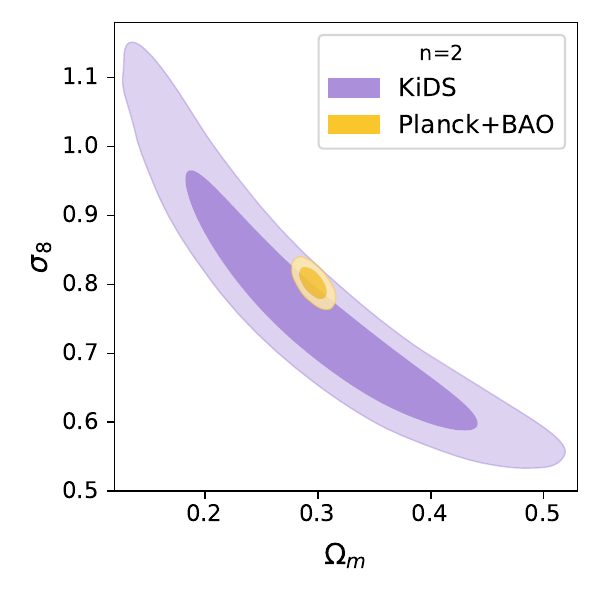}
    \includegraphics[width=.48\textwidth]{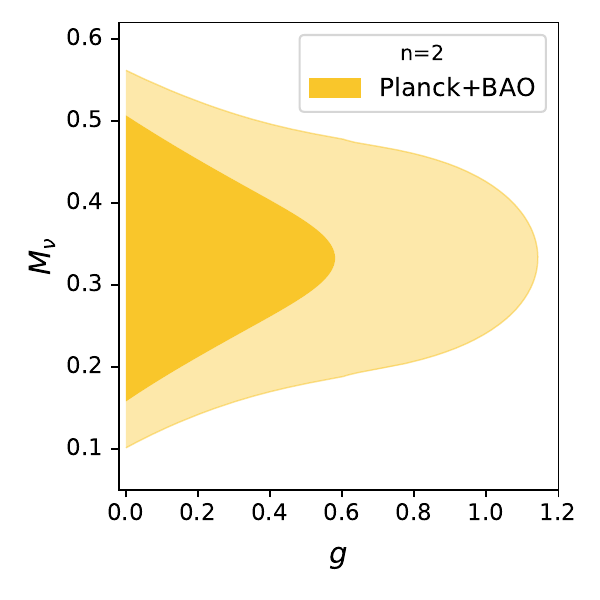}  
    \caption{The same contour confidence plots as in figure \ref{fig:n1_1}, for the $n=2$ KGB model. }
    \label{fig:n1_2}
\end{figure}

\begin{figure}
    \centering
    \includegraphics[width=.48\textwidth]{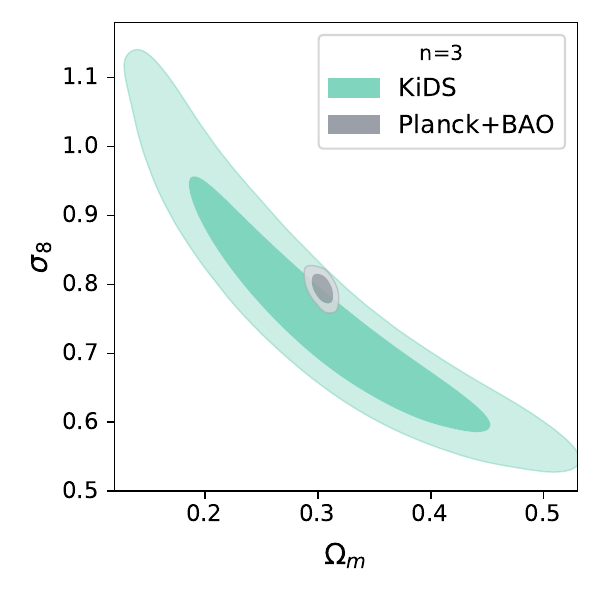}
    \includegraphics[width=.48\textwidth]{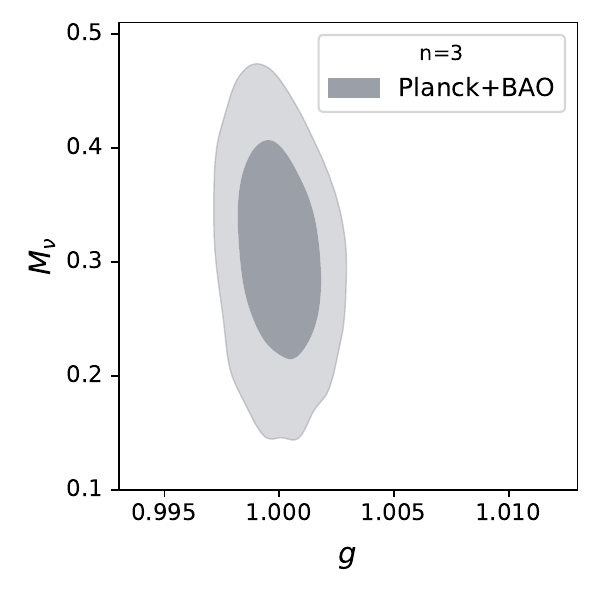}  
    \caption{The same contour confidence plots as in figure \ref{fig:n1_1}, for the $n=3$ KGB model.}
    \label{fig:n1_3}
\end{figure}

It is known that in $\Lambda$CDM an increasing neutrino mass causes a decreasing value of the Hubble constant, and hence it worsens the tension with the distance-ladder determination \cite{Riess:2020fzl}. Some MG models, such as Galileons, work in the other direction, ameliorating this tension \cite{Barreira:2014jha}. Comparing the $H_0$ best fit results from table \ref{table1} for Planck+BAO data with the reported value \cite{Riess:2020fzl} $H_0 =73.2\pm 1.3 \, \rm{km (s Mpc)^{-1}}$ we found that the tension is reduced for these $n$ values in comparison to $\Lambda$CDM.  In fact, there is about $0.43 \sigma$ ($n=1$), $1.72 \sigma$ ($n=2$), and  $2.39 \sigma$ ($n=3$) tension, from which we conclude that the $n=1$ model shows no tension with distance-ladder values, but bigger $n$'s increase the  tension. When a $n$KGB  is in its attractor solution, the asymptotic equation of state for the scalar field is in the phantom domain ($w_{\phi}<-1$)  during the matter era, and tends to -1 in the scalar field dominated era \cite{Kimura:2010di}. This is reason why this model solves the Hubble tension, as known for phantom Universes \cite{Vagnozzi:2019ezj,Planck:2018vyg}. Note that the KiDS data does not sufficiently constrain the Hubble constant, similar as in $\Lambda$CDM model \cite{Kohlinger:2017sxk}. Another aspect is that   this model predicts growing gravitational potentials at low redshift that can be tested using ISW data cross correlated with clustering  \cite{Ferraro:2014msa}.  This has been done in 
\cite{Renk:2017rzu}, where a conflict to fit data is shown in this model.  Another property of general  $n$KGB models is that they exclude the null neutrino mass, see right panels of figures \ref{fig:n1_1}, \ref{fig:n1_2},  and \ref{fig:n1_3}. These results are in agreement with current neutrino mass constraints since all the confidence contours (95\% CL.) for this parameter are larger than the  lower limit of 60 meV from neutrino oscillations \cite{ParticleDataGroup:2020ssz} and are below the upper bound $M_{\nu} < 0.8$ eV (90 \% CL.) from the tritium decay in KATRIN \cite{Aker:2021gma}. If the mass scale happens to be measured closer to the lower bound, small $n$ models could be in trouble to explain this small mass. A normal neutrino hierarchy would put stringer limits than an inverted one.

Another important aspect is to check how the so-called $\sigma_8 $ tension behaves in $n$KGB models; note however that this tension may have a different origin in the $\Lambda$CDM model, see e.g. \cite{Nunes:2021ipq}. It has been previously reported that the $n=1$ model tends to rise the value of $\sigma_8$ \cite{Barreira:2014jha}, but the effect of massive neutrinos is to lower it. The results when these two modifications are considered at once are shown in  table \ref{table1}, where we observe no tension for the considered $n$ values, using Planck+BAO and KiDS data. In small $n$ models there is no $\sigma_8$ tension. This effect  is also evident in the left panels of figures \ref{fig:n1_1}, \ref{fig:n1_2},  and \ref{fig:n1_3}, where we observe that the values from our Planck+BAO selection are compatible at 1$\sigma$ with the results from the KiDS collaboration.  The results in this work are self consistent when considering both datasets and the same $n$ value, and also are consistent with $\Lambda$CDM results from  Planck 2015 \cite{Hildebrandt:2016iqg} and from Planck 2018 \cite{Planck:2018vyg} collaborations. 

To quantify how viable the considered $n$KGB models are compared to $\Lambda$CDM we use the Akaike Information Criterion (AIC)\cite{AIC, Burnham}; recall that a model with a $\Delta$AIC bigger than 10 is essentially not supported, and values approaching to zero are strongly supported. Our results in table \ref{table1} indicate that models with $n=1,2,3$ are supported by KiDS data, however $n=1$ is discarded by  Planck+BAO data; $n=2, 3$ values represent models that, while are not best supported, are not discarded.

\begin{center}
\begin{table*} 
\caption{Constraints at 95\% confidence level on the cosmological and $n$KGB parameter models with $n=1,2,3$ and $M_{\nu}$ $[eV]$ as free parameter. To compare the results with those from $\Lambda$CDM we  include the $\Delta$AIC values defined as $\Delta \rm{AIC}= \rm{AIC}_{nKGB}-\rm{AIC}_{\Lambda \rm{CDM}}$. }
\label{table1}
\begin{tabularx}{0.99\textwidth} { 
   >{\centering\arraybackslash}X 
   >{\centering\arraybackslash}X 
   >{\centering\arraybackslash}X
   >{\centering\arraybackslash}X
   >{\centering\arraybackslash}X
   >{\centering\arraybackslash}X
   >{\centering\arraybackslash}X
   }&\multicolumn{2}{c}{$n=1$}&\multicolumn{2}{c}{$n=2$}&\multicolumn{2}{c}{$n=3$}\\ \hline
&Planck+BAO &  KiDS & Planck+BAO &  KiDS &  Planck+BAO  &  KiDS \\
\hline  
 Parameter &  95\% limits& 95\% limits& 95\% limits& 95\% limits& 95\% limits& 95\% limits\\
\hline 
{\boldmath$100~\omega_b       $} & $2.223^{+0.026}_{-0.026}   $&$2.23^{+3.6}_{-3.4}$& $2.240^{+0.027}_{-0.026}   $& $2.23^{+0.37}_{-0.34}$&$2.246^{+0.017}_{-0.016}   $&$2.23^{+0.36}_{-0.34}$ 
\\ [1ex]

{\boldmath$\omega{}_{c }$} &$0.1204^{+0.0019}_{-0.0019}$& $0.132^{+0.080}_{-0.078}$&$0.1185^{+0.0020}_{-0.0019}$ &$0.139^{+0.082}_{-0.081}$ & $0.1181^{+0.0015}_{-0.0017}$&$0.143^{+0.086}_{-0.084}$\\  [1ex]

{\boldmath$H_0            $} & $72.8^{+1.5}_{-1.5}        $&$75.5^{+6.6}_{-9.4}   $&$69.9^{+1.4}_{-1.4}        $&$75.4^{+6.6}_{-9.5}   $&$68.8^{+1.2}_{-1.2}        $&$75.4^{+6.7}_{-9.7}   $\\  [1ex]

{\boldmath$ 10 ^9 A_s            $} & $2.066^{+0.059}_{-0.064}   $&$2.8^{+1.3}_{-1.1}  $& $2.044^{+0.067}_{-0.069}   $&$2.7^{+1.2}_{-1.0}         $&$2.033^{+0.067}_{-0.075}   $&$2.7^{+1.2}_{-1.0}         $\\  [1ex]

{\boldmath$n_s            $} & $0.9604^{+0.0073}_{-0.0073}$&$1.16^{+0.14}_{-0.19}   $&$0.9658^{+0.0075}_{-0.0075}$&$1.15^{+0.15}_{-0.20}      $& $0.9669^{+0.0054}_{-0.0054}$&$1.15^{+0.15}_{-0.20}      $\\  [1ex]

{\boldmath$\tau{}_{reio } $} & $0.048^{+0.015}_{-0.016}   $ &--&$0.045^{+0.017}_{-0.017}   $&--&$0.042^{+0.015}_{-0.017}   $&--\\  [1ex]

{\boldmath$M_{\nu}        $} & $0.46^{+0.13}_{-0.13}   $& $<2.27     $& $0.33^{+0.13}_{-0.13}   $&$<2.33      $&$0.31^{+0.12}_{-0.13}   $&$<2.37     $\\  [1ex]
{\boldmath$g              $} & $1.4^{+2.7}_{-1.2}         $&$3.2^{+4.5}_{-3.2}         $&$0.24^{+0.73}_{-0.26}      $&$12^{+14}_{-12}            $&$0.9999^{+0.0021}_{-0.0021}$&$15^{+17}_{-15}            $\\  [1ex]
{\boldmath$\Omega_{m}     $} & $0.278^{+0.015}_{-0.014}   $&$0.29^{+0.15}_{-0.14}      $&$0.296^{+0.016}_{-0.015}   $&$0.31^{+0.15}_{-0.14}      $&$0.304^{+0.013}_{-0.013}   $&$0.31^{+0.15}_{-0.15}      $\\  [1ex]
{\boldmath$\sigma_{8}     $} & $0.817^{+0.033}_{-0.033}   $&$0.75^{+0.23}_{-0.20}      $&$0.802^{+0.030}_{-0.031}   $&$0.75^{+0.23}_{-0.19}      $&$0.794^{+0.028}_{-0.030}   $&$0.74^{+0.23}_{-0.19}      $\\  [1ex]
{\boldmath $\Delta$AIC}  & $15.4$ & $3.4$& $7.7$& $3.8$& $8.8$ &$3.7$\\
\hline
\end{tabularx}

\end{table*}
\end{center}

In section \ref{MG_models}, we set the conditions for the dynamical system to be in the attractor solution, following eqs.  (\ref{attractor2}, \ref{w-attractor}). This condition in turn implies a specific determination of our constant $g$ in terms of $\Omega_i$ and $H_0$ \cite{Kimura:2010di}. We checked where in our parameter space this solution is, founding that for $n=1$ and $n=2$ KGB models, the attractor solution lies within the 1-$\sigma$ of our resulting parameters; for  $n=3$ the best fit is out of the attractor solution, similar as in other models studied in ref. \cite{Cardona:2020ama}.

Finally, we have performed a fit to the $n$KGB models letting both $n$ and $g$, together with the neutrino mass, to be free parameters. We use Planck+BAO data and assumed a prior $0.5 \le n \le 30$, the lowest  limit stemming from the physical requirement to avoid perturbation instabilities and the upper limit was found to be big enough, after performing a few runs. We found the $n$ is bounded from above at 95\% CL. by $22.5$. The contours plots are depicted in figure \ref{fig:n_free} and the best fits for the other cosmological parameters at 95\% CL.  are: $100\omega_b=2.247^{+0.029}_{-0.030} $, $\omega_c=0.1174^{+0.0023}_{-0.0023}$, $H_0=68.5^{+1.8}_{-1.8} $, $10^9A_s=2.030^{+0.081}_{-0.10}$, $n_s=0.9688^{+0.0078}_{-0.0082}$, $\tau_{reio}=0.042^{+0.021}_{-0.027}$, $M_{\nu}=0.27^{+0.15}_{-0.14} $, $g<16.6$, $n<22.5$, $\Omega_m=0.305^{+0.019}_{-0.018}$, $\sigma_8=0.794^{+0.028}_{-0.030}$. The $\Delta \rm{AIC}$ for this model is 10.2. Its likelihood is a little smaller than the cases in which $n$ is a fixed parameter, but the model is penalized by the extra parameter, so $n$ as a free parameter is strongly disfavored. When comparing its neutrino mass with the one obtained in $\Lambda$CDM for the same data set, given by  $M_\nu <0.12$ within 95\% CL. \cite{Vagnozzi:2017ovm,Planck:2018vyg}, this model results in a different interval that points to higher masses.

\begin{figure} 
    \centering
    \includegraphics[width=.94\textwidth]{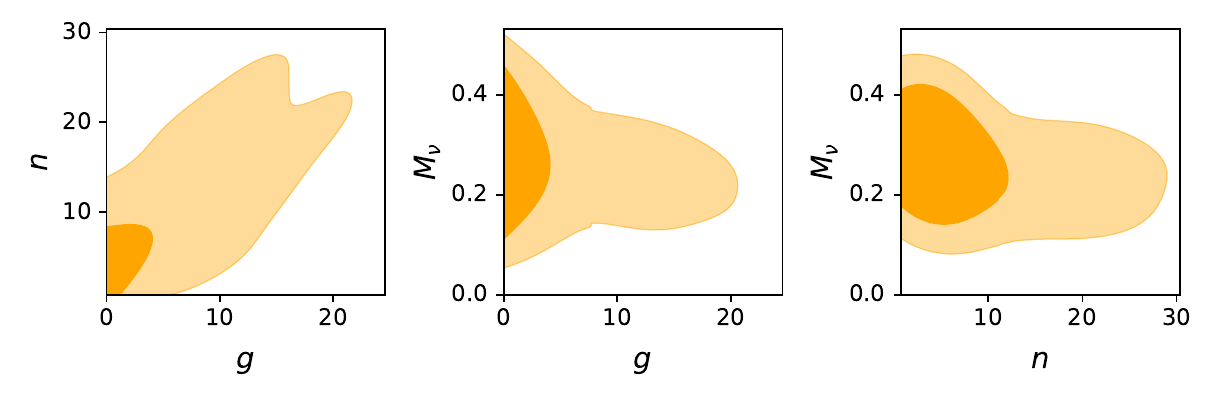}
    \caption{Contour confidence plots using Planck+BAO data for $n , g$, and $M_{\nu}$ at 68 and 95\% CL. In these plots $n$ was considered as a free parameter. }
    \label{fig:n_free}
\end{figure}

\end{section}

\begin{section}{Conclusions} \label{conclu} 
There is an increasing interest in the scientific community to test gravity at large scales. In this line of thought one considers models beyond GR, that in turn imply a larger parameter space.  We investigated the cosmological effects of braiding, $n$KGB models, that possess two extra two parameters $(n,g)$ and added effects of massive neutrinos to study possible parameter degeneracies.

We first consider the CMB TT power spectrum, founding deviations  from $\Lambda$CDM results mainly at large angles, that are difficult to discriminate due to large statistical uncertainties of the cosmic variance; however, cross correlating CMB temperature fluctuations with clustering is an interesting way to test these theories, as shown in refs. \cite{Ferraro:2014msa,Renk:2017rzu,Kable:2021yws}. 

The fifth force associated to the scalar field induces an extra gravitational pull that changes the growth history at late times, and this affects the matter PS. We find that in braiding cosmologies the effect is to enhance PS in a way that already at linear scales it is possible to distinguish $n$KGB models from GR. Even in the case that one increases neutrino masses in $n$KGB models, to lower the PS, this changes in a way that is not possible to mimic the GR PS with lighter neutrinos masses, as it is evident in figures \ref{fig:mps} and \ref{fig:mps2}. This distinction is not possible to achieve in HS gravity for the level of precision required, as shown in refs. \cite{Bellomo:2016xhl,Hagstotz:2019gsv}. It is possible however to compute the effect of peculiar velocities in the PS that have shown to add features in MG two-point observables $P_{\delta \theta}$, $P_{\theta \theta}$ \cite{Hagstotz:2019gsv}, or RSD  \cite{Wright:2019qhf,Garcia-Farieta:2019hal}. The key ingredient that distinguishes these statistics is the growth rate, that for MG models depend on both scale and redshift. We have computed $f(k,z)$ and added neutrino masses to compare its effects. We found that the effects of neutrino masses are smaller than the gravitational effect due to the fifth force of $n$KGB models. In general, the effects of MG are more evident at lower redshifts, since the growth function has more time to rise at scales where MG imprints special features.   We found that at large scales (small $k$) the $f(k,z)$ value tends to the $\Lambda$CDM value but at smaller, still linear scales, the growth rate is larger for $n$KGB models. At $z=1$,  it grows around 4\% between $k=10^{-3}$ and $k=10^{-2}$ $h$/Mpc for the $n=3$ model and 8\% for $n=1$.
At $z=0$, the changes are bigger, it grows around 35\% between $k=10^{-3}$ and $k=10^{-2}$ $h$/Mpc for the $n=3$ model and 70\% for $n=1$. This behavior is expected since the fifth force, that enhances the gravitational pull, increases the growth of structures, as well.  Smaller  deviations of the growth rate are reported for the HS model \cite{Li:2015poa} at $z=0$. For completeness we also computed $P_{\delta \theta}$ and $P_{\theta \theta}$ and compared our results to $\Lambda$CDM model, where the differences to the later model are evident. The most important effect is in $P_{\theta \theta}$ since it depends quadratically on $f(k,z)$ in comparison to $P_{\delta \theta}$ that depends on it linearly, as can be appreciated in figure \ref{fig:croos_power}. Overall the effect of neutrino mass in $f(k,z)$ is much less important than the  $n$KGB gravitation.     

We constrained the base cosmological parameters, the sum of neutrino masses as well as the parameters of $n$KGB models. We fitted the models to our selected Planck+BAO data, and apart to KiDS correlation function data. Best fitted values are displayed in table \ref{table1}, showing that: i) the n=1 KGB model reports no $H_0$ tension using Planck+BAO data, whereas bigger $n$ tends to $\Lambda$CDM and therefore the tension exists. The reason is that KGB models have a phantom behavior and therefore, as known, these models tend to fix this tension \cite{Planck:2018vyg}; ii) All considered $n$KGB models show no $\sigma_8$-tension, when employing  Planck+BAO and KiDS data. The reason for this is that although $n$KGB models increase the PS, as $n$ is bigger, the best fitted values of neutrino mass go to lower values, that help to relax the tension.    

Though some of these models have nice properties such as alleviating the $H_0$ and $\sigma_8$ tensions, all predict a bump in the linear matter PS, as seen in figures \ref{fig:mps} and \ref{fig:mps2}, that is also enhanced through $f(k,z)$, and that surpasses by a significant factor the results of $\Lambda$CDM, cf. plots \ref{fig:croos_power}.   Since we know from present data that possible deviations of power spectra are less than few percent from that of $\Lambda$CDM, see e.g. \cite{Kobayashi:2021oud}., then $n$KGB models are virtually excluded as realistic MG models. 

\end{section}

\acknowledgments
 The authors wish to thank A. Aviles for discussing some of the results of this work. The authors acknowledge partial support by Conacyt project 283151 and SNI. U.N and G.G.A are grateful to FORDECYT PRONACES-CONACYT for support of the present research under Grant CF-MG-2558591 and CF-140630-UNAM-UMSNH.  U.N. thanks the Programa para el Desarrollo Profesional Docente of the Secretar\'{\i}a de Educaci\'{o}n P\'{u}blica (PRODEP-SEP) of M\'{e}xico and the Coordinaci\'{o}n de la Investigaci\'{o}n Cient\'{\i}fica of the Universidad Michoacana de San Nicol\'{a}s de Hidalgo (CIC-UMSNH).

\bibliographystyle{JHEP}  

\bibliography{references.bib}  

\end{document}